# Article information

## Article title
*Clip-on lens for scanning tunneling luminescence microscopy*

## Authors
*Aleš Cahlík[1,*,$], Cinja C. Müller[1], Fabian D. Natterer[1,*]*

## Affiliations
[1]*Department of Physics, University of Zurich, Winterthurerstrasse 190, CH-8057 Zurich, Switzerland*
[$]*Present Address: Department of Applied Physics, Aalto University, Aalto, FI-00076 Finland*

## Corresponding author's email address and Twitter handle
[*]*cahlikales@gmail.com*
[*]*fabian.natterer@physik.uzh.ch*

## Abstract
We demonstrate and verify the in-situ addition of a collecting lens for electroluminescence experiments to an existing scanning tunneling microscope. We fabricate a simple clip-on lens that we reversibly attach at the sample plate via regular sample transfer tools to collimate the light emitted from a plasmonic tunneling junction to the viewport ordinarily used for optical access. The proximity of the lens to the tunneling junction allows us to exploit the full numerical aperture which helps us achieve good collection efficiencies, demonstrating the quick turnaround of converting an existing setup with optical access into a practical scanning luminescence microscope. We verify the function of the clip-on lens by measuring the bias dependent plasmon of Au, Ag, and spatial luminescence maps.

## Specifications table

| Subject area | Physics and Astronomy |
|---|---|
| **More specific subject area** | *Scanning tunneling luminescence microscopy* |
| **Name of your method** | *Clip-on lens for scanning luminescence microscopy* |
| **Name and reference of original method** | *NA* |
| **Resource availability** | *STEP File of lens holder* |

## Background

The next best thing to a purpose-built scanning tunneling luminescence microscope (STLM), is to upgrade an existing scanning tunneling microscope (STM) and to install a light collection system. Unfortunately, the latter is not always an option, may it be because of budget constraints or pushbacks due to technical difficulties associated with installing a permanent lens, space constraints, or the possible irreversibility of such a modification. In that case, the clip-on lens that we describe in this work might be a viable alternative that can be used in-situ and reversibly enhance existing STM for luminescence experiments, provided they have direct optical access to the tunneling junction; a quite common setup for many STM to coarse align the tip before launching the auto-approach towards tunneling contact.

In the present work, we use a variable temperature STM from Omicron (VTSTM) to present our clip-on lens concept that can be shuttled via load lock and transfer stages to the STM head. We use the wobble stick to reversibly attach the lens onto the sample holder in the STM position and successfully verify its photon collection performance using the plasmons of Au(111) and Ag(111). Due to its versatility, the concept may easily be transferable to other STM setups, including low temperature systems.

## Method details
### Clip-on lens design considerations:

The primary objective of the optical system in STLM is to maximize the capture of photons emitted from the tunneling junction and transfer them to ambient detectors. The first elements of a typical setup, installed closest to the junction, are lenses [1–3], parabolic mirrors [4–8], or optical fibers [8,9] (see also reviews in Refs. [10,11]). Unlike purpose-built systems that achieve high photon capture rates with optimal placement of these elements, upgrading an existing STM can present challenges due to space limitations or technical issues with modifying the SPM head and usually requires rather a long service time. For our goal of straightforward in-situ modification, using fiber is intrinsically impractical, and mirrors typically take up too much space near the junction.

We instead use a lens as the collecting element (Figure 1a). The focal point of the lens should coincide with the point-like tunneling junction and have its optical axis ideally aligned at ~30° (Figure 1b), which is the angle of maximal emission [12]. A lens that is placed on this axis will accordingly capture the fraction of photons corresponding to the azimuth of the lens' opening angle. In purpose-built systems, a single or several lenses with large numerical apertures (NA) are usually chosen to encompass the largest possible solid angle. This allows their placement tens of millimeters away from the tunneling junction to not interfere with sample transfer or adatom/molecule deposition. Since our clip-on lens is added after sample transfer, we can place a smaller lens only a few millimeters away from the junction and still exploit its full NA.

At this stage, the particularities of one's setup need to be considered. One important factor is the availability and layout of coarse positioners that will influence how the focal point of the lens is brought into coincidence with the tunneling junction. In our system, we coarse-move the tip in X/Y to the lens' focal point, which we have fixed at 2 mm above the sample plate, equivalent to the typical thickness of our crystals (Figure 1b). Looking at our VTSTM, we also require the lens to be transferable from the load lock to the STM using the existing wobble stick compatible with the standard flag style sample holders and we want to maintain the possibility for high-temperature sample preparation without an attached lens, and for in-situ evaporation onto the sample in the STM. In our case, the VTSTM has several axes with optical access to the tunneling junction that are coming at an angle of about 30° from the sample, which is close to the optimal emission angle and that defines the optical axis of the system. We design an aluminum holder that places a commercial aspheric lens (Thorlabs 355397, Ø7.2 mm, $f$ = 11 mm, working distance = 9.3 mm, NA = 0.3) in the center of this axis about 9 mm away from the center of a prospective sample which is also the presumed lens' focal point (Figure 1a,b). We fix the lens in the respective opening of the holder by a few drops of UHV compatible glue on the side of the aperture. Following the criteria listed by Keizer [13], the opening angle of ~34.7° covers a solid angle of 0.285 sr, which is equivalent to a hemispherical coverage of about 4.5%. Considering the narrow emission profile of a plasmonic junction [12], our configuration leads to a maximum ~9% of the solid angle from which luminescence emission can be received.

For the transfer through the vacuum system, we equip the clip-on lens with a handle and a groove (Figure 1c) that can be grabbed with the wobble stick and that receives the handle of a flag style sample plate (Figure 1a). In the same way, handles of blank or other sample plates in the system are used to shuttle the clip-on lens across the stages from the load lock to the manipulator, to the sample storage, and ultimately to the sample plate loaded in the STM head.

**Clip-on lens validation:**

We have machined an ultra-high-vacuum compatible version of the clip-on lens from two pieces of aluminum (Figure 1c). Figure 2 shows the STM head without and with a clip-on lens attached to the sample plate. To verify the optical path, we require at least one more optical element to collimate the parallelized light from the clip-on lens into our detector. Since the clip-on lens is the only optical element inside the vacuum chamber, we accordingly build the optical setup on the ambient side using commercially available off-the-shelf components, notably a 30/60 mm cage system clamped to the DN40CF viewport feedthrough (Thorlabs VFA275A) and also seen Figure 2d. We correct for the parallel shift of the window's optical axis with respect to the lens' optical axis using a translator stage (Thorlabs LCPX1/M). To couple the parallelized light into the detector, we use an aspheric lens (Thorlabs AC254-030-AB, Ø25.4 mm, $f$ = 30 mm) that focuses the light into a multimode fiber (Thorlabs M133L02, Ø200 µm). The other side of the fiber is connected to a single photon detector (MPD SPAD PD-050-CTE) whose photon arrival clicks are registered with an event recorder (Swabian Instruments, Time Tagger Ultra). The aspheric coupling lens is mounted on a Z-translation stage (Thorlabs SM1ZA) and the fiber coupler (Thorlabs SM05FC) on an XYZ-translation stage (Thorlabs CXYZ05A/M).

The alignment of the ambient side's lens is done by temporarily using the optical system in reverse. We place a fiber coupled laser (Fiberpoint 250MD) instead of the photon detector and change the relative fiber-lens distance until the fiber opening is in the focal point of the coupling lens, which we verify by checking for constant laser spot size along the optical axis using a white screen or projecting the laser spot onto the sample and lens holder. The latter can be useful to first "walk" and center the parallelized laser into the clip-on lens from where it focusses the parallelized laser light onto the sample, producing a visible spot (Figure 1d). We use the XY and Z stages to focus the laser spot on the sample. Using the coarse positioners of the VTSTM, we next move the tip to the center of the laser spot (Figure 1d) and then auto-approach to get into tunneling contact. This configuration represents our basis for further optimization when the tip is brought into the tunneling range during luminescence verification.

For luminescence experiments, we use Au(111) and Ag(111) crystals and Au or Ag coated Pt/Ir tips created by indenting the tip into the respective substrate. While the VTSTM may be cooled, we conduct our measurements at room temperature and at a pressure of about $5 \times 10^{-9}$ mbar. At a current setpoint of 45 nA and bias of 3.6 V, we record a photon flux of up to $2 \times 10^5$ cps in our initial tests on Au(111). Considering typical quantum efficiencies (number of emitted photons per electron) of plasmonic junctions between $1 \times 10^{-3}$ and $1 \times 10^{-4}$ [14], the overall collection efficiency of the optical system ranges between 0.7% and 7%, respectively. These latter numbers include all losses occurring along the optical path after entering the clip-on lens, including windows, fiber-coupling, and they also contain the quantum efficiency of the single photon detector ($QE_{max}$ = 0.49 at 550 nm). Our collection yield is respectable compared to purpose built systems [13] and is sufficient to venture deeper into STLM experiments or to gain first-light experience.

We further verify the functioning of the clip-on lens for an Ag(111) crystal mounted on a different sample plate. We cleaned the crystal only by sputtering Ag with 1.5 keV Ar$^+$ ions for 45 minutes. When we measure the bias voltage dependence of the plasmon on an Ag(111) terrace between 1 to 7 V (Figure 3a) at a constant current of 45 nA, we clearly see the about 500-fold increase of the luminescence signal with a maximum photon count around 4 eV, in good agreement with the literature [6,15,16].

To show the utility of the clip-on lens also for spatial luminescence mapping, we select a region containing different surface terminations (Figure 3b) because we found the variations of the luminescence yield across Ag steps rather small for our sample and measurement conditions. The stepped circular holes in the Ag layer are due to the absent annealing after sputtering. Albeit of unknown origin, the amorphous layer shows an overall brighter luminosity signal than the Ag termination (Figure 3c) and the spatially distributed photon counts clearly trace the boundary between the adlayer and the Ag termination as well as the granular structures within the adlayer. To measure the location dependent luminescence, we use a constant bias of 3.6 V at 5 nA and synchronize the photon counting with the motion of the tip by collecting all photons between consecutive locations using a pixel trigger as the start/end point for the event-logger.

**Discussion and outlook**

Although we built the lens holder from aluminum for vacuum experiments and to verify the transfer from the load lock to the sample holder, we note that our clip-on lens does not necessarily require ultra-high vacuum compatibility but solely low outgassing because we can remove the lens from the system during bakeout. This could be an interesting feature because it would allow 3D printing of polymer lens-holders, the use of high refractive index optical components made from polymers, and of coated optics to optimize the optical properties of the system.

The use of our clip-on lens concept also inherently resolves problems associated with fixed lens systems that only have coarse motion of the tip in the Z direction while X and Y coarse motions are realized on the sample stage. There, any deviation of the tip apex from the focal point during an experiment cannot be in-situ corrected and likely requires a tip-transfer while the clip-on lens on the sample stage would allow shifting the apex back into the actual focal point.

Our clip-on lens is an accessible addition to our existing STM system, which can inspire the upgrade of other machines. It enables new spatial measurement modalities that were previously not available. The addition of an optical collection system with appreciable efficiency and the quickly gathered expertise could justify spectral analysis tools as the logical next step in one's research.

**List of used components:**

- Lens holder machined from two aluminum pieces
- 2×M1.6 stainless steel mounting screws
- Vacuum lens: Thorlabs 355397
- Glue to fix lens, Epotek E20H
- Viewport clamp for cage system: Thorlabs VFA275A
- 6mm rods, 12 pcs
- Cage adapter: Thorlabs TTN247583
- Cage translator: Thorlabs LCPX1/M
- Z-translation stage: Thorlabs SM1ZA
- Ambient side aspheric coupling lens: Thorlabs AC254-030-AB
- XYZ translation stage: Thorlabs CXYZ05A/M
- Fiber coupler: Thorlabs SM05FC
- Multimode fiber: Thorlabs M133L02
- Single Photon detector: MPD SPAD PD-050-CTE
- Alignment laser: Fiberpoint 250MD
- Photon counter: Swabian Instruments - Time Tagger Ultra

**CRediT author statement**
*Aleš Cahlík: Conceptualization, Methodology, Supervision, Validation, Investigation, Funding acquisition, Writing - Review & Editing. **Cinja C. Mueller**: Investigation. **Fabian D. Natterer**: Conceptualization, Validation, Investigation, Writing – original draft, Supervision, Funding acquisition.*

**Acknowledgments**
*We thank the technical workshop of the Department of Physics at the University of Zurich for their expert assistance and K.-H. Ernst for providing us with the variable temperature STM. We thank Anna Rosławska, Katharina Kaiser, and Martin Švec for fruitful discussions and advice.*
*Funding: This work was supported by the Swiss National Science Foundation [200021_200639, PP00P2_211014, 206021_213238, CRSK-2_221052].*

**Declaration of interests**
☒ The authors declare that they have no known competing financial interests or personal relationships that could have appeared to influence the work reported in this paper.

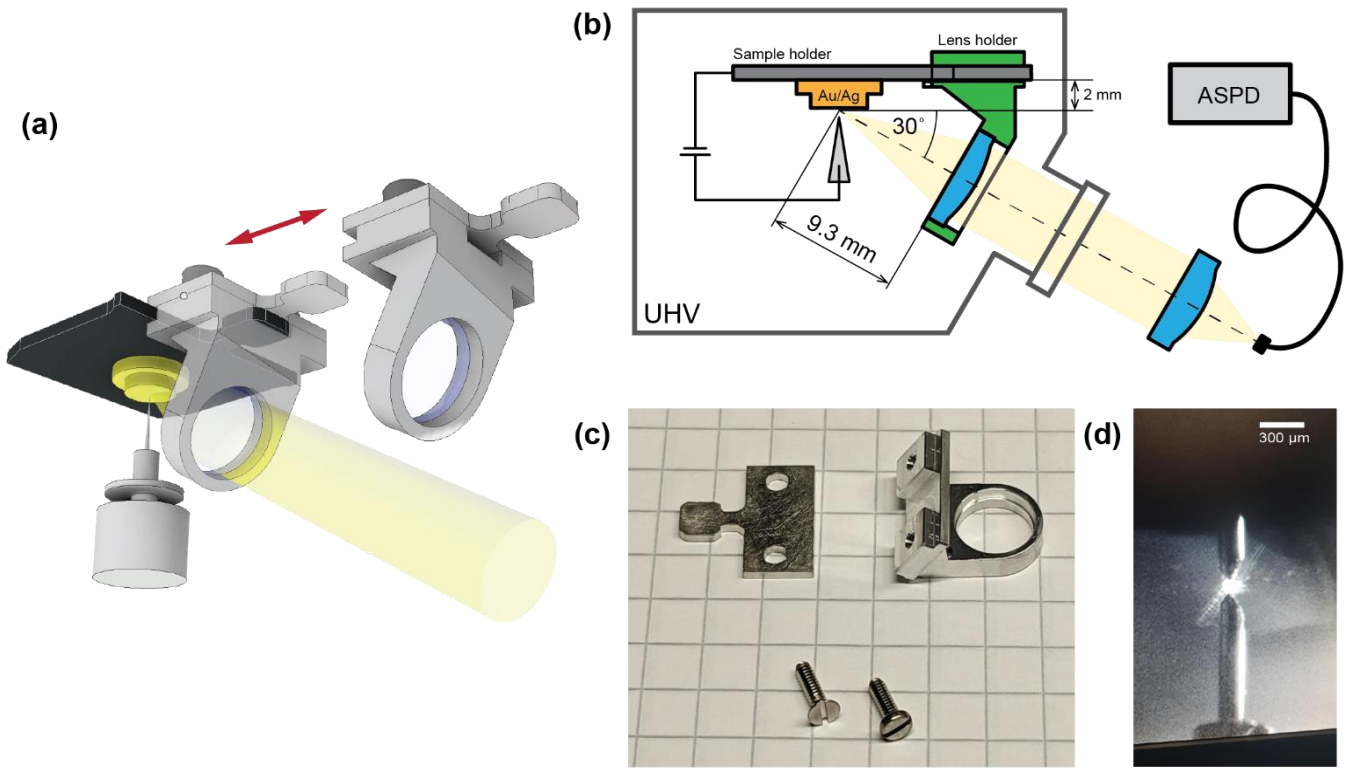

*Figure 1:* **Clip-on lens concept and alignment for scanning luminescence microscopy**. *(a) Lens holder that can be reversibly mounted on the handle of a sample plate, allowing for small focal distances between the tip-location and the lens to exploit the full numerical aperture of the lens. (b) Optical path along the optimal luminescence angle of 30° from the tunneling junction via the clip-on lens (green) to the outside of the vacuum chamber, ultimately feeding via focusing lens into a fiber coupled avalanche single photon detector (ASPD). (c) The machined clip-on lens components with the hole accepting the sample handle shown on 5 mm grid paper. The lens is held by the groove and fixed with a dab of epoxy adhesive (Epotek E20H). (d) The tip and its mirror image coarse walked into the focal point of the lens, illuminated via a laser pointer through the lens' optical path.*

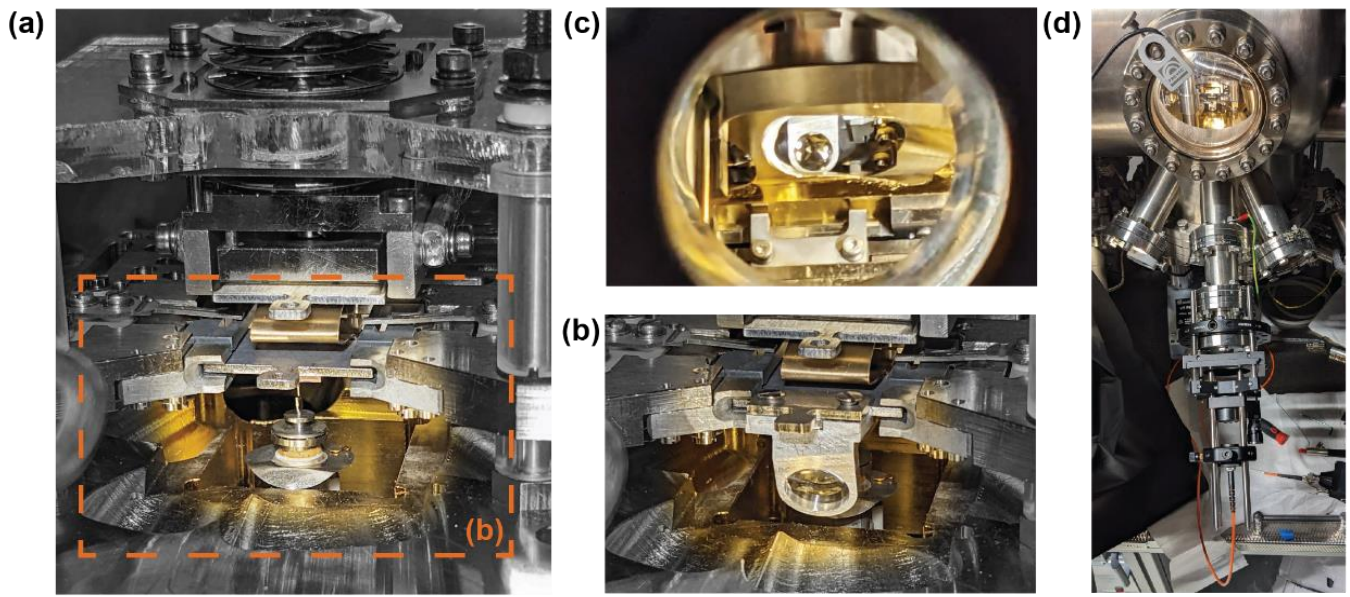

Figure 2: **Clip-on lens transfer into STM.** (a) View of the STM head before attaching the clip-on lens and (b) after attaching the lens to the handle of the sample plate using the regular wobble stick. (c) View along the optical axis viewport with mounted clip-on lens. (d) View of ambient side optical setup showing the flange mount cage system, translator stages, coupling lens, and optical fiber.

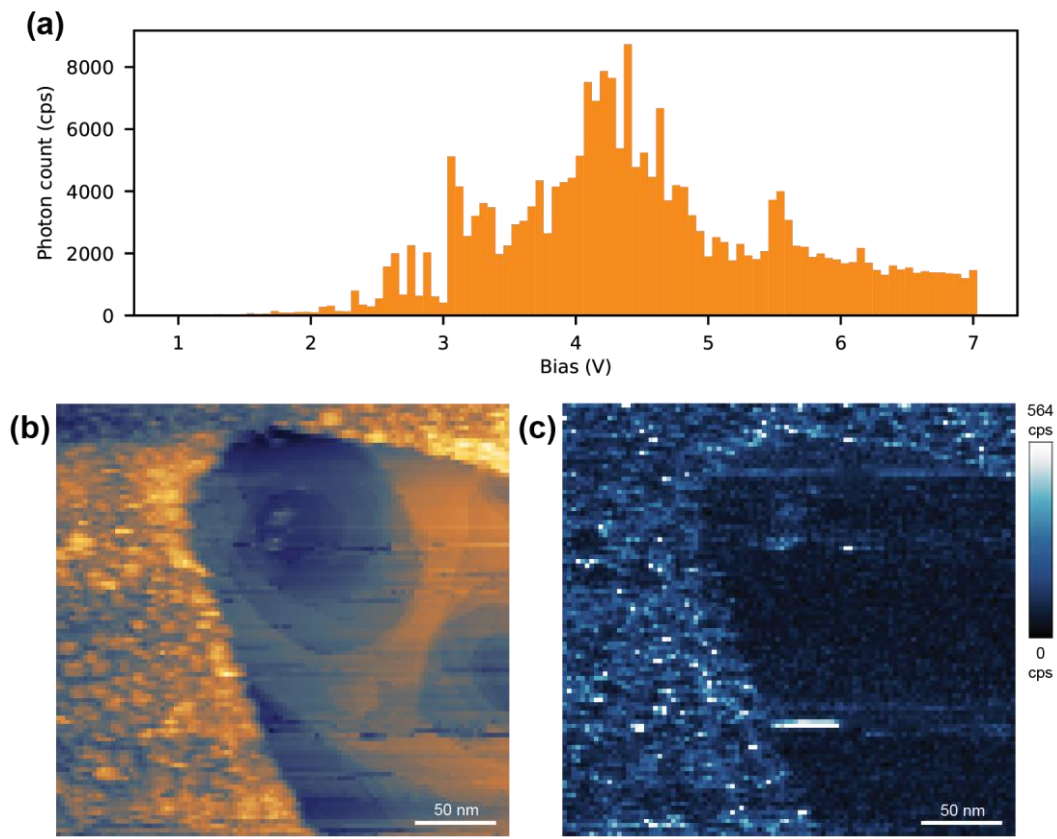

Figure 3: **Luminescence verification using clip-on lens**. (a) Photon count vs bias voltage on Ag(111) (T=300 K, I=45 nA) showing the characteristic plasmonic enhanced emission around about 4 eV. The flicker in the data is due to junction instabilities. (b) Topographic image of an amorphous overlayer and sputter holes on Ag(111) and (c) simultaneously measured luminescence signal showing spatial contrast between the substrate and the amorphous layer as well as within the layer itself (V=3.6 V, I=5 nA, 500 ms dwell time). The topographic image was corrected by a linear line-by-line subtraction and the luminescence map filtered with a 1-pixel Gaussian blur.